\documentclass{eas}
\usepackage{graphicx}
\usepackage[authoryear]{natbib}
\begin{document}

\title{Infrared Properties Of AGB Stars: from existing databases to Antarctic surveys.}
\runningtitle{Guandalini \etal: Infrared Properties Of AGB Stars
\dots}
\author{R. Guandalini}\address{Dipartimento di Fisica, Universit\`a di Perugia,
Via A. Pascoli, 06123 Perugia, Italy;
\email{guandalini@fisica.unipg.it }}
\author{M. Busso}\sameaddress{1}
\begin{abstract}
We present here a study of the Infrared properties of Asymptotic
Giant Branch stars (hereafter AGB) based on existing databases,
mainly from space-borne experiments. Preliminary results about C
and S stars are discussed, focusing on the topics for which future
Infrared surveys from Antarctica will be crucial. This kind of
surveys will help in making more quantitative our knowledge of the
last evolutionary stages of low mass stars, especially for what
concerns luminosities and mass loss.
\end{abstract}
\maketitle
\section{Introduction}

Towards the end of their life, stars of low and intermediate mass
($M < 8 M_{\odot}$) evolve along the Asymptotic Giant Branch (AGB)
phase [see Busso {\em et al.\/} \cite{Busso9} and references
herein for more details]. In this stage they experience extensive
phenomena of mass loss that affect deeply their evolution.
Sedlmayr \cite{sedlmayr} shows that stellar winds are also
fundamental for the enrichment of the Interstellar Medium that is
replenished by these stars with about 70\% of all the matter
returned after stellar evolution. Moreover, AGB stars provide the
starting conditions for the formation of planetary nebulae.

The radiative emission of the cool dust in the infrared (IR)
normally dominates the energy distribution of AGB stars
(particularly for the most evolved ones) and this fact is mainly
due to their strong stellar winds. Until recently the bolometric
magnitude of the most evolved AGB stars was difficult to derive,
due to insufficient photometric coverage of the mid-IR range of
the electromagnetic spectrum (the importance of mid-IR
observations for AGB stars is clearly shown in Guandalini {\em et
al.\/} \cite{guandalinia}, Figure 1). The availability of large IR
databases from space-borne telescopes like ISO, IRTS, MSX has
substantially improved this situation. At the same time, Hipparcos
distances for AGB stars have been corrected from various biases in
works like the one from Bergeat \& Chevallier \cite{bergeat} and
the period-luminosity relations found for Miras have been
drastically improved as shown in Whitelock {\em et al.\/}
\cite{whitelock}. The study of fundamental physical parameters of
these stars (like luminosity, IR colors and mass loss) can now be
be performed in a rather quantitative way.

However, IR space-borne observations of AGB stars present some
disadvantages. In particular, the duration of the operational
period of the telescope is quite limited, observations with a long
time of integration are difficult and AGB stars are generally
observed at a single epoch. All these facts hinder our
understanding of several basic physical parameters, which are
fundamental in the study of AGB stars. In this respect a
complementary role in solving these problems could be played by
ground-based observations at IR wavelengths, especially from
Antarctica.

The Antarctic Plateau (in particular Dome C) presents the best
conditions available on Earth from the point of view of infrared
observations as shown in several other contributions from this
conference: therefore, Antarctica is the best place where
ground-based observations in the (mid-)IR can be performed. An
Antarctic IR Observatory might be crucial for clarifying the final
stages of stellar evolution by:
\begin{itemize}
    \item observing properties of evolved stars in the Magellanic
    Clouds, at known distance and metallicity different from the Milky Way;
    \item doing the same for the Galactic Center, where the metal
    blend is different and the extreme O-enhancement prevents the
    formation of C stars as shown in Uttenthaler {\em et al.\/} \cite{utten};
    \item looking for mass loss calibrations in the IR.
\end{itemize}

Our efforts are addressed to the preparation for IR observations
of AGB stars from Antarctica through the IRAIT telescope,
presented by Tosti {\em et al.\/} \cite{tosti} in this conference.
With this aim we are trying to understand which kind of
observations from Antarctica is the best and most promising from
the point of view of AGB stars. In Guandalini {\em et al.\/}
\cite{guandalinia},\cite{guandalinib} we are making for these
sources an overview of the IR data available from existing
catalogues that will be completed in future works. In this note we
address some interesting issues concerning AGB stellar variability
from an IR point of view, and the relevance of Antarctic
observations for them.

\section{Infrared Variability}

\begin{figure}[!!h]
\begin{center}
\includegraphics[scale=0.59]{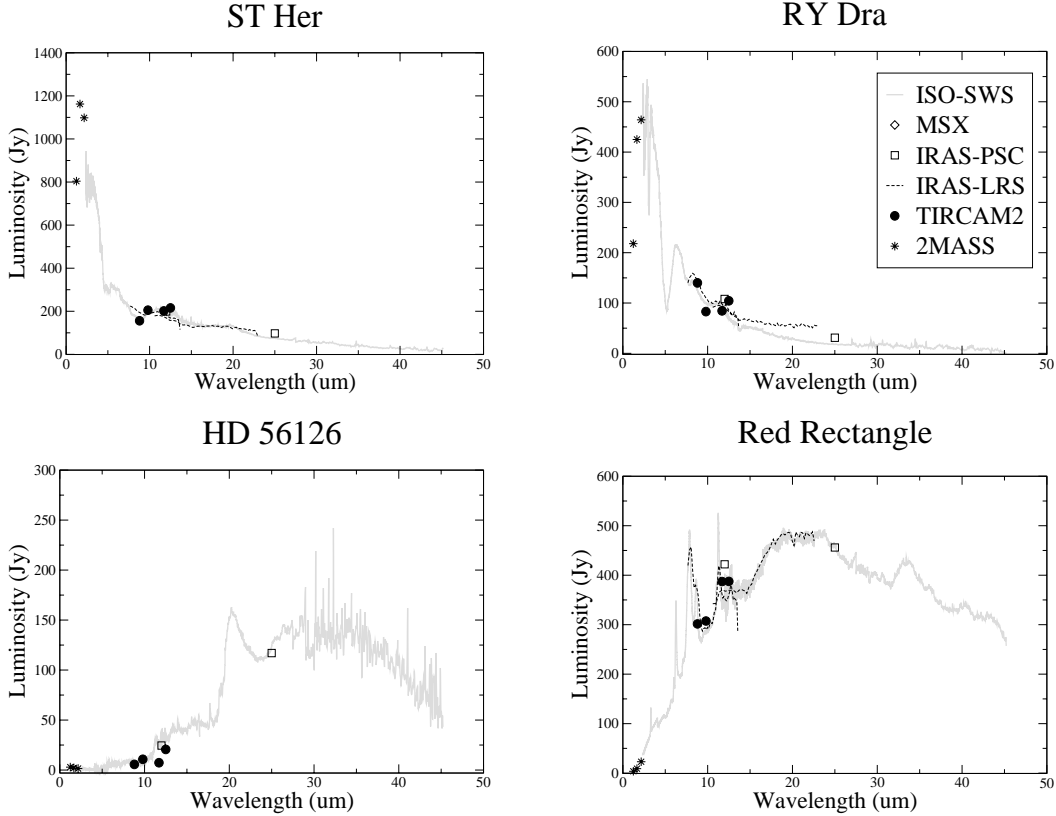}
\caption{The Spectral Energy Distributions (SEDs) of a few
sources, as available from IRAS PSC, IRAS LRS, ISO-SWS, MSX,
TIRCAM2 and 2MASS. SEDs with maximum emission in near-IR, as well
as with maximum emission longward of 20$\mu$m all show a constant
flux in mid-IR.} \label{fig1}
\end{center}
\end{figure}

\begin{figure}[!!h]
\begin{center}
\includegraphics[scale=0.59]{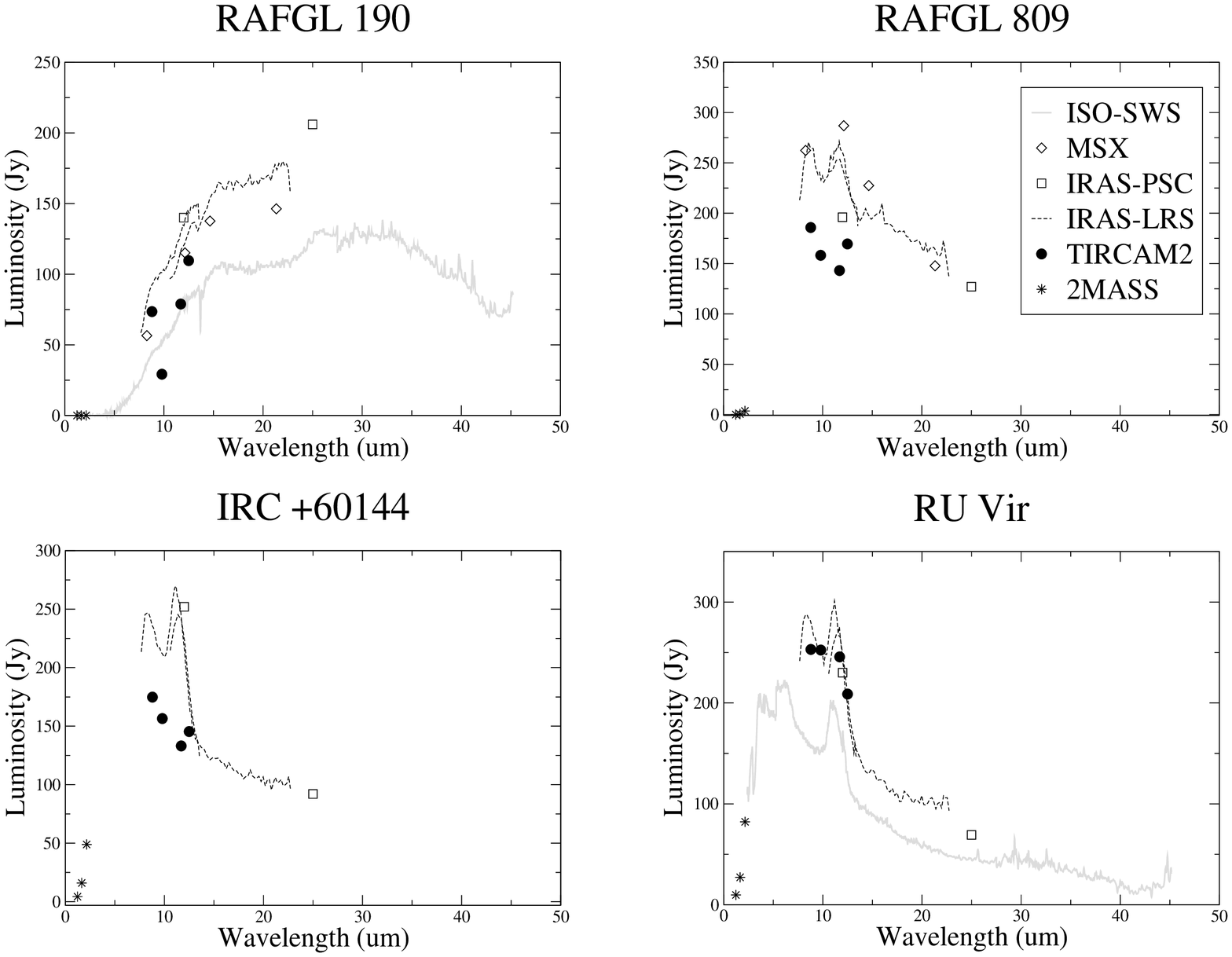}
\caption{SEDs of sources that show significant variability over
the time elapsed from the IRAS to the TIRCAM2 observations. Data
available from IRAS PSC, IRAS LRS, ISO-SWS, MSX, TIRCAM2 and 2MASS
are included. Only sources with maximum emission in the range
$8-20 \mu$m appear to be variable.} \label{fig2}
\end{center}
\end{figure}

AGB stars present strong variability at optical wavelengths and
are roughly divided into three subclasses according to their
variability type: Miras, Semiregulars and Irregulars. Effects due
to variability at IR wavelengths are expected to be less relevant
when compared with the optical ones.

\begin{figure}[!!h]
\begin{center}
\includegraphics[scale=0.65]{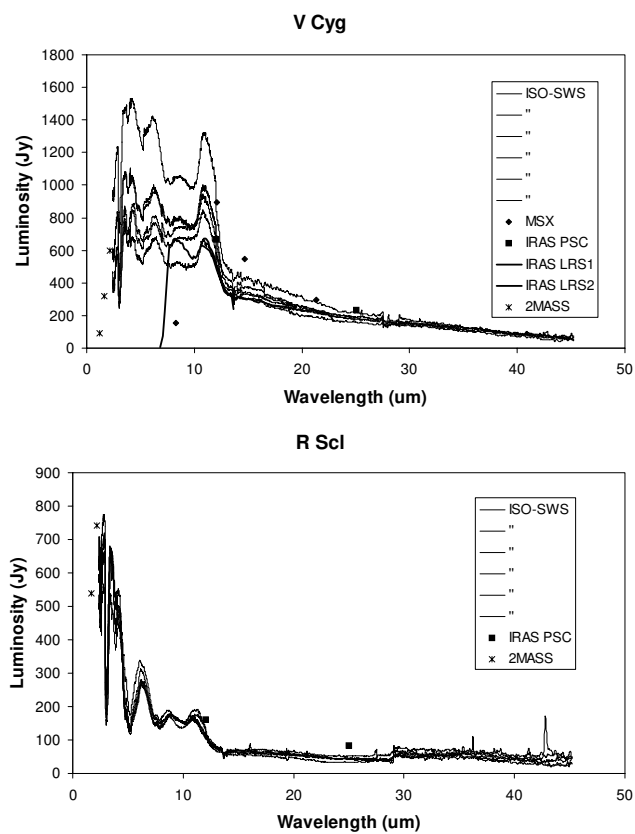}
\caption{SEDs of two sources observed several times with ISO-SWS.
Only the source with maximum emission in the $10 \mu$m region
appears to be variable.} \label{fig3}
\end{center}
\end{figure}

Figures \ref{fig1} and \ref{fig2} present the available
information on the IR SEDs for two groups of AGB sources that are
discussed in detail in Busso {\em et al.\/} \cite{busso6}. Figure
\ref{fig1} shows distributions that share the property of being
non-variable at IR wavelengths over an elapse of time of almost 20
years. They include Semiregular sources with minimal IR excess, in
which the SED is peaked in near-IR. They also include evolved
(post-AGB) objects in which the maximum emission is at very long
wavelengths (from 20 to more than 40 $\mu$m). Instead, Figure
\ref{fig2} shows a group of Mira variables, in which the emission
peaks near 10 $\mu$m: they present the common property of a
long-term mid-IR variability that seems to be restricted to this
special class of sources. Moreover, Figure \ref{fig3} shows two of
the few available AGB sources observed several times by ISO-SWS
and confirms the same behavior: Mira variables present an IR
variability even over a relatively short time interval.

There is not yet an agreement on the origin and properties of this
phenomenon. It needs to be examined in more detail to understand
its nature. It could be perhaps a variability induced by shock
waves caused by dynamic events in the photosphere or by
magneto-hydrodynamical modes (and magnetic storms). Otherwise, it
could be a "simpler" mid-IR variability, originating in the
emission of the circumstellar envelopes and caused by modulations
in the efficiency of dust formation. This would be expected to be
more typical of the Mira sources; Semiregulars have thinner
circumstellar envelopes, while Post-AGB stars should have detached
shells not strongly influenced by the variability of the central
star.

The best way to examine this variability is that of observing AGB
stars at different epochs also in mid-IR and this task can be
performed by a ground-based telescope placed in Antarctica.
Moreover, simultaneous observations in the near- and mid-IR and
correlations with optical variability could be fundamental to
understand these phenomena.

\section{Conclusions: Why Surveys from Dome C}

Important IR studies for AGB stars can be made in the best way
from ground-based locations like Dome C and Antarctica. The study
of the main features regarding IR variability (and also optical
variability) is one of them. It can be fulfilled with, surveys
through wide field (in the future) or small area (IRAIT) imaging
of interesting stellar systems, multiple observations of chosen
AGB sources at different epochs and wavelengths and observation of
AGB sources from Magellanic Clouds, whose distance is
well-estimated.

In this way we could obtain:\begin{itemize}
    \item light curves and therefore good knowledge of
    variability;
    \item an accurate study of the luminosity
variations over a wide region of the electromagnetic spectrum,
including optical and IR;
    \item finally, a reliable comparison between
observations and models of AGB stars.
\end{itemize}



\begin{thebibliography}{99}
\bibitem[2005]{bergeat}
    Bergeat, J., \& Chevallier, L. 2005, A\&A, 429, 235
\bibitem[1999]{Busso9}
    Busso, M., Gallino, R., \& Wasserburg, G. J. 1999, ARA\&A, 37, 239
\bibitem[2006]{busso6}
    Busso, M., Guandalini, R., Persi, P., Corcione, L., \& Ferrari-Toniolo, M. 2006, AJ,
    submitted
\bibitem[2006]{guandalinia}
    Guandalini, R., Busso, M., Ciprini, S., Silvestro, G., \& Persi, P. 2006, A\&A, 445, 1069
\bibitem[2007]{guandalinib}
    Guandalini, R., Busso, M., \& Cardinali, M. 2007, in
    preparation
\bibitem[1994]{sedlmayr}
    Sedlmayr, E. 1994, Lecture Notes in Physics, 428, 163
\bibitem[2003]{stran3}
    Straniero, O., Dom\'{i}nguez, I., Cristallo, S., \& Gallino, R. 2003, PASA 20, 389
\bibitem[2007]{tosti}
    Tosti, G. \etal\ 2007, this conference
\bibitem[2006]{utten}
    Uttenthaler, S., Hron, J., Lebzelter, T., Busso, M., Schultheis, M., \& Kaeufl, H. U. 2006,
    astro-ph/0610500
\bibitem[2006]{whitelock}
    Whitelock, P. A., Feast, M. W., Marang, F., \& Groenewegen, M. A. T. 2006,
    MNRAS, 369, 751
\end{thebibliography}
\end{document}